\documentclass[prd,showpacs,preprintnumbers,nofootinbib,aps]{revtex4}
\usepackage{subeqnarray}
\usepackage{epsfig,amsmath,amssymb}
\usepackage{mathrsfs}
\usepackage[usenames,dvipsnames]{color}
\usepackage[pagebackref=false, colorlinks=false]{hyperref}
\definecolor{redish}{rgb}{0.7,0.2,0.0}  % color defined in (r=red,g=green,b=blue) model
\definecolor{bluish}{rgb}{0.2,0.5,0.8}
\hypersetup{linkcolor=redish,          % color of internal links
                  citecolor=blue,        % color of links to bibliography
                  filecolor=magenta,      % color of file links
                  urlcolor=bluish}          % color of the url
\DeclareFontFamily{U}{rsfs}{}         % Formal Script            %
\DeclareFontShape{U}{rsfs}{m}{n}{<5> rsfs5 <6><7> rsfs7          %
  <8><9><10><10.95><12><14.4><17.28><20.74><24.88> rsfs10}{}     %
\DeclareMathAlphabet{\mathfs}{U}{rsfs}{m}{n}                     %

\newcommand{\ba}{\nopagebreak[3]\begin{eqnarray}}
\newcommand{\ea}{\end{eqnarray}}
\newcommand{\bii}{\begin{itemize}}
\newcommand{\eii}{\end{itemize}}

\begin{document}

\title{Thermal Stability of Black Holes With Arbitrary Hairs}
\author{Aloke Kumar Sinha}
\email{akshooghly@gmail.com}
\affiliation{ Haldia Government College,West Bengal, India\\
Ramakrishna Mission Vivekananda Educational and Research Institute, Belur Math, India}
\pacs{04.70.-s, 04.70.Dy}
       
\begin{abstract}

We have derived the Criteria for thermal stability of charged rotating black holes, for horizon areas that are large relative to the Planck area (in these dimensions). In this paper, we generalised it for black holes with arbitrary hairs.The derivation uses results of loop quantum gravity and equilibrium statistical mechanics of the Grand Canonical ensemble and there is no explicit use of classical spacetime geometry at all in this analysis. The assumption is that  the mass of the black hole is a function of its horizon area and all the hairs. Our stability criteria are then tested in detail against some specific black holes, whose metrics provide us with explicit relations for the dependence of the mass on the area and other hairs of the black holes. This enables us to predict which of these black holes are expected to be thermally unstable under Hawking radiation.

\end{abstract}
\maketitle

\section{Introduction}

The semiclassical analysis of thermal instability of nonextremal, asymptotically flat black holes with negative specific heat \cite{dav77} motivates to study the same,  entirely from quantum perspective like Loop Quantum Gravity \cite{rov,thie}. A consistent understanding of quantum black hole entropy can be obtained from Loop Quantum Gravity \cite{abck98,abk00}.The modified Bekenstein-Hawking area law has been derived \cite{km98}-\cite{bkm10} for macroscopic (astrophysical) black holes, with the leading correction being logarithmic in area with the coefficient $-3/2$.\\
 
 Classically a black hole in general relativity is characterized by its mass ($M$), charge ($Q$) and angular momentum ($J$).  Intuitively, therefore, we expect that thermal behaviour of a quantum black hole will depend on all of its hairs. These additional hairs will change the energy function of the black hole. This will in turn make the stability criteria more complicated. It is obvious that stability criteria will now depend on the interplay among all hairs . So, it is really interesting to redo, what we have done earlier \cite{aspm16}, taking the contributions of additional hairs. We will do this in this paper in details. We will assume that the mass of black hole is a function of the horizon area and its all hairs.\\ 

 The idea of thermal holography \cite{pm07}, \cite{pm09}, and the saddle point approximation are used to evaluate the canonical partition function corresponding to the horizon, retaining Gaussian thermal fluctuations. This leads to a general criterion of thermal stability as a set of inequalities connecting 1st and 2nd order area derivatives of microcanonical entropy along with 1st and 2nd order derivatives of mass with respect to area and all the hairs. These are derived in detail in this paper. This inequality is nontrivial when the microcanonical entropy has corrections (of a particular algebraic sign) beyond the area law, as is the case for the loop quantum gravity calculation of the microcanonical entropy \cite{km00}. This procedure has recently been done for charged rotating black hole{\cite{aspm16} and  the derived stability criterion indeed `predicts' the thermal instability of asymptotically flat Kerr-Newman black holes contrasted with the thermal stability of anti-de Sitter Kerr-Newman black holes (for a range of its parameters). \\

The paper is organized as follows: In section 2, the idea of thermal holography, alongwith the concept of (holographic) mass associated with horizon of a black hole is briefly reviewed and the grand canonical entropy of a large quantum black hole, with arbitrary number of hairs , is determined. In section 3, the criterion for thermal stability of such black holes is determined by using saddle point approximation to evaluate the horizon partition function for Gaussian thermal fluctuations around thermal equilibrium. Here, we also compare our stability criteria with that obtained earlier as a particular case. In the next section, this stability criterion is used to test on various black holes, with the objective of predicting their behaviour under decay due to Hawking radiation. We end in section 5 with a brief summary and outlook.

\vspace{.3 in}
\section{Thermal holography}

 In this section, we present a generalization of the thermal holography for rotating electrically charged quantum radiant horizons discussed in \cite{aspm16}, to the situation when the horizon has arbitrary number of hairs. To make this section self-contained, some overlap with \cite{aspm16} is inevitable.

\subsection{Mass Associated With horizon}

Black holes at equilibrium are represented by isolated horizons, which are internal boundaries of spacetime. Hamiltonian evolution of this spacetime gives the first law associated with isolated horizon($b$) and is given as,
\begin{eqnarray}
\delta E^{t}_{h}=\frac{\kappa^{t}}{8\pi}\delta A_{h}+ P_{i}^{t}\delta C^{i}_{h}
\end{eqnarray}
Here, Einstein summation convention is used  i.e.  summation over repeated indice $i$ from $1$ to $n$(=total number of hairs) is implied
and $E^{t}_{h}$ is the energy function associated with  the horizon. $\kappa^{t}$, $P_{i}^{t}$  are respectively the surface gravity associated with the area of horizon($A_{h}$) and the potential corresponding to the charge (hair) $C^{i}_{h}$. The label '$t$' denotes the particular time evolution field ($t^{\mu}$)  associated with the spatial hypersurface chosen. $E^{t}_{h}$ is  assumed here to be a function of $A_{h}$ and all $C^{i}_{h}$.\\

 The advantage of the isolated (and also the radiant or {\it dynamical}) horizon description is that one can associate with it a mass $M_h^t$, related to the ADM energy of the spacetime through the relation
\begin{eqnarray}
E_{ADM}^t = M_h^t + E_{rad}^t ~\label{adm}
\end{eqnarray}
where, $E^{t}_{rad}$ is the energy associated with spacetime  between the horizon and asymptopia. An isolated horizon admits $E^t_{rad} \neq 0$, and hence a mass is defined {\it locally} on the horizon.  

\subsection{Quantum Geometry}

The Hilbert space of  a generic quantum spacetime is given as, $\mathcal{H}=\mathcal{H}_{b}{\otimes}\mathcal{H}_{v}$ , where $b(v)$ denotes the boundary (bulk)  space. A generic quantum state is  thus given as
\begin{equation}
\vert\Psi\rangle=\sum\limits_{b,v} C_{b,v} \vert\chi_{b}\rangle {\otimes} \vert\psi_{v}\rangle ~\label{genstate} 
\end{equation} 
Now, the full Hamiltonian operator ($\widehat{H}$),  operating on $\mathcal{H}$ is given by
\begin{equation}\label{hamil}
\widehat{H}\vert\Psi\rangle=(\widehat{H_{b}}{\otimes}I_{v}+I_{b}{\otimes}\widehat{H_{v}})\vert\Psi\rangle
\end{equation} 
where,  respectively, $I_{b} (I_{v})$ are identity operators on $\mathcal{H}_{b} (\mathcal{H}_{v})$ and $\widehat{H_{b}} (\widehat{H_{v}})$ are the Hamiltonian operators on $\mathcal{H}_{b}(\mathcal{H}_{v})$. \\

 The first class constraints are realized on Hilbert space as annihilation constraints on physical states. The bulk Hamiltonian operator thus annihilates bulk physical states
\begin{eqnarray}
\widehat{H_{v}}\vert\psi_{v}\rangle=0 \label{bulkham}
\end{eqnarray}

 The bulk quantum spacetime is assumed to be free of any charge(hair), so that eqn. (\ref{bulkham}) is augmented by the relation
\begin{eqnarray}
[\widehat{H_{v}} - P_{i} \widehat{C^{i}_{v}}] | \psi_{v} \rangle = 0 ~.~\label{fullham}
\end{eqnarray}

\subsection{Grand Canonical Partition Function}

Consider the black hole immersed in a heat bath, at some (inverse) temperature $\beta$, with which it can exchange energy, charge, angular momentum and all quantum hairs. The grand canonical partition function of the black hole is given as,
\begin{eqnarray}
Z_{G}=Tr(exp(-\beta\widehat{H}+\beta P_{i}\widehat{C^{i}})) ~\label{gcpf}
\end{eqnarray}
where  the trace is taken over all states.  This definition,  together with eqn.s (\ref{genstate}) , (\ref{bulkham}) and (\ref{fullham}) yield
\begin{eqnarray}
Z_{G} &=& \sum_{b,v} \vert C_{b,v} \vert^{2} \langle\psi_{v}\vert\psi_{v}\rangle \langle\chi_{b}\vert  exp(-\beta\widehat{H_{b}}+\beta P_{i}\widehat{C^{i}})  \vert\chi_{b}\rangle \nonumber \\ 
&=& \sum\limits_{b} \vert C_{b} \vert^{2} \langle\chi_{b}\vert exp(-\beta\widehat{H_{b}}+\beta P_{i}\widehat{C^{i}})  \vert\chi_{b}\rangle ~,~ \label{pf+ham}
\end{eqnarray}
assuming that the bulk states are normalized. The partition function  thus turns out to be  completely determined by the boundary states ($Z_{Gb}$), i.e.,
\begin{eqnarray}
Z=Z_{Gb} &=& Tr_{b} \exp(-\beta\widehat{H}+\beta P_{i}\widehat{C^{i}}) \nonumber \\
&=& \sum\limits_{l,k_{1},...,k_{n}} g(l,k_{1},...,k_{n}) \hspace{.1 in} \exp \Big(-\beta \Big(E(A_{l},C^{1}_{k_{1}},...,C^{n}_{k_{n}})-\sum_{i=1}^{n}  P_{i}C^{i}_{k_{i}} \Big)\Big) ~.~\label{bdypf}
\end{eqnarray}
Where $g(l,k_{1},...,k_{n})$ is the degeneracy corresponding to energy $E(A_{l},C^{1}_{k_{1}},...,C^{n}_{k_{n}})$ and $l,k_{i}$ are the quantum numbers corresponding to area and charge $C^{i}$ respectively.  Here,  the spectrum of  the boundary Hamiltonian operator is assumed to be  a function of area and all other charges of the boundary, considered here to be the horizon. Following \cite{mm12}, it is further assumed that these 'hairs' all have a discrete spectrum. In the semiclassical limit of quantum isolated horizons of macroscopic area, they all have 
large eigenvalues i.e. ($l,k_{i}>>1$), so that, application of the Poisson resummation formula \cite{cm04} gives
\begin{equation}
Z_{G}=\int dx\hspace{.05 in} \Big(\prod\limits_{i=1}^{n}\int dy_{i}\Big) \hspace{.05 in} g(A(x),C^{1}(y_{1}),...,C^{n}(y_{n}))\hspace{.05 in}  \exp \Big(-\beta \Big(E(A(x),C^{1}(y_{1}),...,C^{n}(y_{n}))-\sum_{i=1}^{n}  P_{i}C^{i}({y_{i}})\Big)\Big)
\end{equation}
where $x,y_{i}$ are respectively the continuum limit of $l,k_{i}$ respectively. 

Following \cite{mm12}, we now assume that the semiclassical spectrum of the area and charges  are linear in their arguments, so that a change of variables gives, with constant Jacobian, the result
\begin{equation}
Z_{G}=\int dA\hspace{.05 in} \Big(\prod\limits_{i=1}^{n}\int dC^{i}\Big) \hspace{.05 in}   \exp \Big(S(A)-\beta \Big(E(A,C^{1},...,C^{n})-  P_{i}C^{i}\Big)\Big)~,~ \label{pfresult}
\end{equation}
where, following \cite{ll}, the {\it microcanonical} entropy of the horizon is defined by $\exp S(A) \equiv \frac{ g(A(x),C(y_{1}),...,C(y_{n}))}{\frac{dA}{dx}\frac{dC^{1}}{dy_{1}}...\frac{dC^{n}}{dy_{n}}}$.

\section{Stability Against gaussian Fluctuations}

\subsection{Saddle Point Approximation}

 The equilibrium configuration of black hole is given by the saddle point $\bar{A},\bar{C^{i}}$ in the $(n+1)$ dimensional space of integration over area and $n$ charges. This configuration is identified with with an  isolated horizon, as already mentioned. The idea now is to examine the grand canonical partition function for fluctuations $a=(A-\bar{A}), c^{i}=(C^{i}-\bar{C^{i}})$ around the saddle point, in order to determine the stability of the equilibrium isolated horizon under Hawking radiation. We restrict our attention to Gaussian fluctuations. Taylor expanding eqn (\ref{pfresult}) about the saddle point, yields 
\begin{eqnarray}
Z_{G} &=& \exp[ S(\bar{A})-\beta M(\bar{A},\bar{C^{1}},...,\bar{C^{n}})+\beta P_{i}\bar{C^{i}}] \nonumber \\
&\times &\int dA\hspace{.05 in} \Big(\prod\limits_{i=1}^{n}\int dC^{i}\Big)  \exp \{-\frac{1}{2}[( \beta M_{AA}-S_{AA} )a^{2} +2 \sum_{i=1}^{n} \beta M_{AC^{i}}ac^{i} \nonumber \\
&+&  \sum_{i=1}^{n}\sum_{j=1}^{n} \beta M_{C^{i}C^{j}}c^{i}c^{j}] \} ~,~ \label{sadpt}
\end{eqnarray}
where  $M(\bar{A},\bar{C^{1}},...,\bar{C^{n}})$ is the mass of equilibrium isolated horizon. Here $ M_{AC^{i}} \equiv \partial^2 M/\partial A \partial C^{i} \vert_{(\bar{A},\bar{C^{1}},...,\bar{C^{n}})}$ etc. 

Observe that all observables of Loop Quantum Gravity used here are self-adjoint operators over the boundary Hilbert space, and hence their eigenvalues are real. It suffices therefore to restrict integrations over the spectra of these operators to the real axes.

Now,  in the Saddle point approximation the coefficients of  terms linear in $a,c^{i}$ vanish by definition of the saddle point. These imply that, at saddle point
\begin{equation}\label{beta}
\beta=\frac{S_{A}}{M_{A}} ,\hspace{.2 in} P_{i}=M_{C^{i}} 
\end{equation}

\subsection{Stability Criteria}

Convergence of the integral (\ref{sadpt}) implies  that the Hessian matrix ($H$) has to be positive definite, where
\begin{eqnarray}
 H = \left( \begin{array}{ccccc}
 \beta M_{AA}- S_{AA} \hspace{.3 in}& \beta M_{AC^{1}} \hspace{.3 in} & \hspace{.3 in}\beta M_{AC^{2}} \hspace{.3 in} & ......... \hspace{.3 in}& \beta M_{AC^{n}} \vspace{.1 in} \\  
\beta M_{AC^{1}} \hspace{.3 in}& \beta M_{C^{1}C^{1}} \hspace{.3 in}& \beta M_{C^{1}C^{2}} & ......... \hspace{.3 in}& \beta M_{C^{1}C^{n}} \vspace{.1 in} \\
\beta M_{AC^{2}} \hspace{.3 in}& \beta M_{C^{2}C^{1}} \hspace{.3 in}& \beta M_{C^{2}C^{2}} & ......... \hspace{.3 in}& \beta M_{C^{2}C^{n}} \vspace{.1 in} \\ 
....... \hspace{.3 in}& ....... \hspace{.3 in}& ...... & ......... \hspace{.3 in}& ...... \vspace{.1 in} \\ 
\beta M_{AC^{n}} \hspace{.3 in}& \beta M_{C^{n}C^{1}} \hspace{.3 in}& \beta M_{C^{n}C^{2}} & ......... \hspace{.3 in}& \beta M_{C^{n}C^{n}} \vspace{.1 in} \\ 
 \end{array} \right) \label{hess}
\end{eqnarray}
Here, all the derivatives are calculated at the saddle point.The Hessian matrix is real symmetric and hence it can be diagonalized by orthogonal matrix. So, positive definiteness of Hessian matrix  boils down to the positivity of $(n+1)$ eigenvalues of Hessian matrix. Hence the stability criteria is equivalent to the criteria of positivity of all eigenvalues of Hessian matrix  and is given as :
\begin{eqnarray}
D_{1}>0,D_{2}>0,....,D_{n+1}>0 \label{stab1} 
\end{eqnarray}
Where,
\begin{eqnarray}
 D_{1}=\beta M_{AA}- S_{AA} , \hspace{.3 in} D_{2} = \Bigg\vert \begin{array}{cc}
\beta M_{AA}- S_{AA} \hspace{.3 in}& \beta M_{AC^{1}}  \\  
\beta M_{AC^{1}} \hspace{.3 in}& \beta M_{C^{1}C^{1}}  \\
  \end{array} \Bigg\vert,  \nonumber \\ 
  D_{3}=\Bigg\vert \begin{array}{ccc}
\beta M_{AA}- S_{AA} \hspace{.3 in}& \beta M_{AC^{1}} \hspace{.3 in} & \beta M_{AC^{2}}  \\  
\beta M_{AC^{1}} \hspace{.3 in}& \beta M_{C^{1}C^{1}} \hspace{.3 in}& \beta M_{C^{1}C^{2}}  \\
\beta M_{AC^{2}} \hspace{.3 in}& \beta M_{C^{2}C^{1}} \hspace{.3 in}& \beta M_{C^{2}C^{2}}  \\ 
\end{array} \Bigg\vert ,....., D_{n+1}= \vert H \vert \label{stab2}
\end{eqnarray}
Where, $\vert H \vert =$ determinant of the Hessian matrix $H$. \\
 
Of course, (inverse) temperature $\beta$ is assumed to be positive for a stable configuration.  What is new is the requirement that temperature must increase with horizon area, inherent in the positivity of the quantity ($\beta M_{AA} - S_{AA}$) which appears in several of the stability criteria. If this is violated, as for example in case of the standard Schwarzschild black hole, thermal instability is inevitable.\\

The convexity property of the entropy follows from the condition of convergence of partition function under gaussian fluctuations \cite{cm04}, \cite{ll}, \cite{mon}.  The thermal stability is related to the convexity property of entropy. Hence, the above conditions are correctly the conditions for thermal stability. For rotating charged horizons, eqn.s (\ref{stab1}) and (\ref{stab2})reproduce the thermal stability criterion with $n=2$ i.e. $D_{1}>0,D_{2}>0,D_{3}>0$ with the identification that charge of the black hole($Q$)=$C^{1}$ and angular momentum of the black hole($J$)=$C^{2}$. It can be easily checked that these three conditions correctly reproduce the Earlier\cite{aspm16} seven conditions of thermal stability of charged rotating black holes. Eqn.s (\ref{stab1}) , (\ref{stab2}) necessarily tell us that thermal stability of black hole is a consequence of the interplay among all the charges of the black hole. \\

As claimed in the Introduction, the thermal stability criteria above are derived by the application of standard statistical mechanical formalism to a quantum horizon characterized by various observables having discrete eigenvalue spectra. Thus, no aspect of classical geometry enters the derivation of these criteria. If the mass of the horizon is given in terms of its area and all the charges  of the horizon , then it is possible, on the basis of our stability criteria, to  predict which  black holes will radiate away to extinction, and which ones might find some stability, and for what range of  parameters. This is what is attempted in the next section.

{\section{Predicting Thermal Stability of Black Holes}

 Notice that in the stability criteria derived in the last section, first and second order derivatives of the microcanonical entropy of the horizon at equilibrium play a crucial role, in making some of the criteria non-trivial. Thus, corrections to the microcanonical entropy beyond the Bekenstein-Hawking area law, arising due to quantum spacetime fluctuations, are very significant, because without these, some of the stability criteria might lose their essential physical content. It has been shown that \cite{km00} the microcanonical entropy for  {\it macroscopic} isolated horizons has the form
\begin{eqnarray}
S~&=&~S_{BH} ~-~\frac32 \log S_{BH} +{\cal O}(S_{BH}^{-1}) ~\label{kment} \\
S_{BH} ~&=& ~ \frac{A_h}{4 A_P}~,~A_P \rightarrow {\rm Planck~area} ~. \label{bek}
\end{eqnarray}
In reference \cite{km00} , the result \ref{kment} was derived for black holes in four dimensional spacetime.This is based on a three dimensional SU(2) Chern-Simons theory. Consideration of U(1) Gauge also gives same correction \cite{bkm10}. Since entropy is a physical quantity, it cannot depend on the choice of gauge fixing. We will assume similar correction of entropy for the examples that we will study i.e. leading order entropy correction even in higher dimension due to quantum gravity is logarithmic with negative coefficient. Although it is an asumption, it can be argued heuristically as follows \cite{psm}: Gauss constraints restrict over the availability of phase space and hence degrees of freedom decreases. So, entropy decreases as a consequence of it. Now, without any constraint, entropy of black hole($S$) is given as $S= \frac{A}{4A_{P}}$. So, leading order correction is expected to be logarithmic in area($A$) with negative coefficient. $4+1$ dimensional Lorentz group i.e. $SO(4,1)$ has $10$ generators. Among these, $6$ generators correspond to rotation of 4-dimensional space. So, analogy of $SO(3,1)$ impiles $SU(2)\times SU(2)$ is the covering group of $SO(4)$ with $6$ generators. Each of these $6$ planes can be associated with an $U(1)$ rotation. Hence the coefficient of the $log(A)$ term is expected to be double of $SO(3)$ case i.e. $-(2 \times 3/2) = -3$. Ofcourse, this is heuristic, the exact number should be given by details of possible embeddings of SU(2) in the covering group of SO(4). The upshot is that correction is logarithmic in area with negative coefficient of order one.

\subsection{Uncharged Rotating Black Hole in (4+1) Dimensional Flat Spacetime}

The properties of uncharged Rotating Black Holes in $(N+1)$ dimensional spacetime had been studied in details in \cite{myers}. We extract out the necessary portions required for $(4+1)$ dimensional spacetime. The Mass($M$) of the Black hole is given as,
\begin{equation}\label{bm}
M= \frac{3\pi\mu}{8G}
\end{equation}
where, $\mu$ is mass parameter of $G$ is the newton constant for five dimensional spacetime.\\

Since the spacetime is five dimensional, there will be two directions of roation for the black hole. The rotational parameters($a_{1}, a_{2}$) are given as,
\begin{equation}\label{ba}
a_{1}= \frac{3J_{1}}{2M} , \hspace{.2 in} a_{2}= \frac{3J_{2}}{2M}
\end{equation}
where, $J_{1}, J_{2}$ are the two angular momentums of the black hole.\\

The horizon of the black hole($r_{h}$) is given as,
\begin{equation}\label{bh1}
\Pi_{r_{h}}= \mu r_{h}^{2}
\end{equation}
where,
\begin{equation}\label{bh2}
\Pi_{r}= (r^{2}+ a_{1}^{2})(r^{2}+ a_{2}^{2})
\end{equation}\
Equation No. (\ref{bh1}) and (\ref{bh2}) together give,
\begin{equation}\label{bh3}
2r_{h}^{2}= \mu - a_{1}^{2}- a_{2}^{2}+ \sqrt{(\mu- a_{1}^{2}- a_{2}^{2})^{2}- 4a_{1}^{2}a_{2}^{2}}
\end{equation}\
So, positivity of $r_{h}^{2}$ ensures that $\mu > (a_{1}^{2} + a_{2}^{2})$ and reality of $r_{h}^{2}$ ensures that either $\mu > (a_{1} + a_{2})^{2}$ or $\mu < (a_{1} \sim a_{2})^{2}$. These imply that reality and positivity of $r_{h}^{2}$ necessarily means $\mu > (a_{1} + a_{2})^{2}$. This mathematical inequality is the artifect of the fact that horizon of five dimensional Myers-Perry black hole is formed only if mass dominates over rotation.\\

The surface gravity($\kappa$) of black hole is given as,
\begin{equation}\label{bsg1}
\kappa= \frac{\frac{\partial (\Pi)}{\partial r}- 2\mu r}{2 \mu r^{2}}\bigg\vert_{r_{h}}
\end{equation}\
Equation No. (\ref{bh2}), (\ref{bh3}) and (\ref{bsg1}) together give,
\begin{equation}\label{bsg2}
\kappa= \frac{\sqrt{(\mu- a_{1}^{2}- a_{2}^{2})^{2}- 4a_{1}^{2}a_{2}^{2}}}{\mu r_{h}}
\end{equation}
The horizon area ($A$) of black hole is given as,
\begin{equation}\label{bar}
A=\frac{16\pi G M (1-\frac{a_{1}^{2}}{r_{h}^{2}+ a_{1}^{2}}- \frac{a_{2}^{2}}{r_{h}^{2}+ a_{2}^{2}})}{3 \kappa}
\end{equation}\
Last equation implies that large horizon area($A$) limit means large value of black hole mass($M$) and small value of surface gravity($\kappa$). Equation No. (\ref{bsg2}) implies that small value of $\kappa$ means large value of $\mu$ and $r_{h}$. \\

Equation No. (\ref{bh3}), (\ref{bsg2}) and (\ref{bar}) together give,
\begin{equation}
A=\frac{16\pi G M}{3}.\frac{1}{\sqrt{(\mu- a_{1}^{2}- a_{2}^{2})^{2}- 4a_{1}^{2}a_{2}^{2}}}. \frac{r_{h}^4- a_{1}^{2}a_{2}^{2}}{r_{h}}
\end{equation}\\

Logarithm of last equation in the limit $\mu > (a_{1}+a_{2})^{2}$ gives ,
\begin{equation}
ln(A)= \frac{3}{2}ln(M)- \frac{27\pi}{32GM^{3}}.(J_{1}^{2}+J_{2}^{2})+ {\cal O}(\frac{J^{4}}{M^{6}})
\end{equation}
where, we expressed $\mu, a_{1}, a_{2}$ is expressed in terms of $M, J_{1}, J_{2}$ by relation (\ref{bm}) and (\ref{ba}). Here,  ${\cal O}(\frac{J^{4}}{M^{6}})$ are terms of order $\frac{J_{1}^{4}}{M^{6}}$, $\frac{J_{2}^{4}}{M^{6}}$, $\frac{J_{1}^{2}J_{2}^{2}}{M^{6}}$ etc and we have thrown away irrelevant constant like $ln(G)$ etc.\\

In large horizon area ($A$) limit, it can be easily shown that $D_{1}=(\beta M_{AA}- S_{AA})= - \frac{1}{6A^{4/3}A_{P}}$, in leading order. So, it is negative. This implies thermal instability of five dimensional myers-Perry black holes under hawking radiation.

\subsection{Asymptotically ADS Dyonic Black Holes with Electric and Magnetic Charge}

The properties of Asymptotically ADS Dyonic Black Holes with Electric and magnetic Charge has been studied in details in (\cite{Wiltshire}) . The $4$ dimensional metric for this black hole is given as,\\
\begin{equation}\label{adsmetric}
ds^{2}= -f dt^{2}+ f^{-1} dr^{2}+ R^{2} d\Omega^{2}
\end{equation}
where,
\begin{equation}\label{ads1}
\phi= \frac{\phi_{3}}{r^{3}}+ {\cal O}(r^{-4})
\end{equation}
\begin{equation}\label{ads2}
f= \frac{-\Lambda r^{2}}{3}+ 1- \frac{2M}{r}+ \frac{Q^{2}+P^{2}}{r^{2}}+ \frac{\Lambda \phi_{3}^{2}}{5 r^{4}} + {\cal O}(r^{-5})
\end{equation}
\begin{equation}\label{ads3}
R= r- \frac{3 \phi_{3}^{2}}{20 r^{5}} +{\cal O}(r^{-6})
\end{equation}
\begin{equation}\label{ads4}
\phi_{3}= \frac{g_{0}}{\Lambda} \int\limits_{r_{h}}^{\infty} \frac{dr}{R^{2}} (Q^{2} exp(2g_{0}\phi) - P^{2} exp(-2g_{0}\phi))
\end{equation}
Where, $r_{h}, Q, P$ are the radius of horizon, electric charge and magnetic charge of the black hole respectively. $\Lambda(<0)$ is the cosmological constant , $g_{0}$ is diatonic coupling constant and $\phi$ is the diatonic field.\\

It is clear from equ no. \ref{ads2} that unless diatonic field is too strong, its contribution on mass of black hole of large horizon is necligible. Again, equ no. \ref{ads1}, \ref{ads2} , \ref{ads3} and \ref{ads4} together imply that weak diatonic field limit is possible if $Q^{2}, P^{2} << A$.\\

So, Area of black hole horizon($A$) for weak field limit is given as, $A=4\pi R^{2}(r=r_{h}) \approx 4 \pi r_{h}^{2}$ for large black hole.\\

The black hole horizon is given as a solution of $f(r= r_{h})= 0$. Considering all the above equations , it turns out that
\begin{equation}\label{adsmass1}
M\approx \frac{ A^{3/2}}{48 l^{2} \pi^{3/2}} + \frac{A^{1/2}}{4\pi^{1/2}}+ \frac{\pi^{1/2}(Q^{2}+ P^{2})}{A^{1/2}}
\end{equation}
Where  $\Lambda= -1/ l^{2}$, $l$ is the cosmic length.\\

Retaining the leading terms in the horizon area, as before, Eqn.s (\ref{beta}), (\ref{kment}) and (\ref{adsmass1}) and give the inverse temperature $ (\beta)$ as
\begin{equation}\label{adskntemp}
\beta=  \bigg(\frac{1}{A_{P}} - \frac{6}{A}\bigg)\bigg / \bigg ( \frac{ A^{1/2}}{8 l^{2} \pi^{3/2}} + \frac{1}{2\pi^{1/2}A^{1/2}}- \frac{2\pi^{1/2}(Q^{2}+ P^{2})}{A^{3/2}}    \bigg) 
\end{equation}
Since we are dealing with macroscopic black holes with a large event horizon area($A>>A_{P}$) \cite{aspm16} and hence  $\beta$ is positive in weak field limit i.e. $Q^{2}, P^{2} << A$.  One can also verify that unlike the asymptotically flat case, for anti-de Sitter black holes, the horizon temperature does increase with horizon area.\\ 

To complete the test for thermal stability, condition (\ref{stab1}) has to be checked with $n=2$. With the identification  $ C_{1}=Q$  and  $C_{2}=P $, it can be easily shown from equation no.(\ref{stab1}) and (\ref{stab2}) that $D_{1} = \bigg(\beta \cdot \big( \frac{1}{64 l^{2} \pi^{3/2}A^{1/2}} - \frac{1}{16\pi^{1/2}A^{3/2}} + \frac{3\pi^{1/2}(Q^{2}+ P^{2})}{4A^{3/2}}\big) - \frac{3}{2A^{2}}\bigg)$ and is positive if $A >> l^{2}$.  In this limit ($ A>>A_{P},l^{2},Q^{2},P^{2} $), it can easily be shown that $D_{2}, D_{3}$ are also positive. Thus, the Asymptotically ADS Dyonic large Black Holes with Electric and Magnetic Charge turns out to be thermodynamically stable for a sufficiently large negative asymptotic curvature in weak field limit. 

\subsection{Summary}

We reiterate that our analysis is quite independent of specific classical spacetime geometries, relying as it does on quantum aspects of spacetime. The construction of the partition function used standard formulations of equilibrium statistical mechanics augmented by results from canonical Quantum Gravity, with extra inputs regarding the behaviour of the microcanonical entropy as a function of area {\it beyond the Bekenstein-Hawking area law}, as for instance derived from Loop Quantum Gravity \cite{km00}. However, we emphasize that the results are more general than being restricted to any specific proposal for quantum spacetime geometry, requiring only certain functional dependences on horizon area and other parameters of statistical mechanical quantities like entropy. It also stands to reason that our stability criteria are useful for predicting the thermal behaviour vis-a-vis Hawking radiation for specific astrophysical black holes.

It is also noteworthy that our approach is applicable to black holes with arbitrary `hairs' (charges) in Lorentzian spacetimes with arbitrary number of spatial dimensions. Black holes with Arbitrary hairs had been studied in (\cite{cpw1})-(\cite{cpw2}). In these references, it is more or less shown that energy of black hole in general depends on all its hairs as well. So, it is naturally expected that these hairs will govern thermal stability of black hole. In this sense, this paper covers the entire gamut of black hairs and their role on thermal stability of black hole. \

\end{document}